# Classical model for measurements of an entanglement witness

Brian R. La Cour[*] and E. C. George Sudarshan

*Applied Research Laboratories, The University of Texas at Austin, P.O. Box 8029, Austin, Texas 78713-8029, USA*
(Received 13 July 2014; published 2 September 2015)

We describe a classical model that may serve as an analog for joint and local measurements of an entanglement witness. The analogous experimental procedure and data analysis protocol of the model follow those of a previous experiment to measure an entanglement witness with polarized photons prepared in a mixed state [Phys. Rev. Lett. **91**, 227901 (2003)]. Numerical simulations show excellent agreement with both experimental results and quantum-mechanical predictions. This agreement is made possible by the fact that the model exhibits contextuality due to the postselection of coincident detection events.

## I. INTRODUCTION

Entanglement is widely recognized as an important resource in quantum information processing [1]. While generally regarded as a uniquely quantum feature, aspects of entanglement are found in classical systems as well [2–4]. Recently, a locally deterministic, detector-based method for modeling quantum measurements has been proposed and shown to exhibit analogous quantum behavior [5]. That such a scheme can be devised raises the question of whether a classical model of this sort is capable of exhibiting true entanglement and, if so, how one might go about verifying its presence.

In pursuing this question, it is important to distinguish several different, related properties often associated with entanglement. In a formal, mathematical sense, entanglement may be defined simply as *nonseparability* of the quantum state. As such, this is a property that is trivially satisfied by many analogous classical systems; for example, a superposition of modes in a vibrating rectangular membrane.

Indeed, from the perspective of quantum computing, nonseparability alone often suffices to realize a significant quantum speedup, as is found in Shor's factoring algorithm or Grover's search algorithm [6,7], although examples such as the Deutsch–Jozsa algorithm illustrate that entanglement is not always needed to realize a quantum speedup in efficiency [8].

Of course, entanglement, as originally conceived by Schrödinger, connotes much more [9]. The mathematical property of nonseparability entails, via the Born rule, certain observational consequences regarding correlations between subsystems. Thus, we observe that electrons in a singlet state, say, exhibit perfect anticorrelation in measurements of their respective spins along a given direction. This limited behavior can be observed rather trivially in classical systems as well. For this reason, separable states which exhibit correlations between different subsystem measurements are sometimes referred to as *classically correlated*.

The Born interpretation of, say, the singlet spin state is generally taken to be independent of the spatial separation between the constituent subsystems. This suggests a certain nonlocal character to the correlations appearing between such systems even though, as noted above, such correlations can arise classically as well. When one considers measurements along different, nonorthogonal directions, however, the naive classical interpretation breaks down. By using this idea, Bell was able to derive an inequality relating measurable correlations between different subsystem observables which, if violated, would imply entanglement [10]. If, furthermore, the subsystem measurements are performed in a spacelike-separated manner, such a violation could not be interpreted classically; a property which is generally referred to as *quantum nonlocality*.

A violation of Bell's inequality clearly indicates entanglement; however, an entangled state may still fail to violate the inequality. To more finely distinguish the transition from separable to entangled states, one may employ an entanglement witness $W$, an operator chosen for a particular class of quantum states and such that, for $\rho$ in that class, $\mathrm{Tr}[W\rho] < 0$ if and only if $\rho$ is entangled [11]. In general, entanglement witnesses may be difficult to define and measure. For certain classes of quantum states, a parsimonious set of projection operations may suffice to construct $W$ [12].

In particular, the class of two-qubit Werner states [13], defined by

$$\rho_q = q|\psi_s\rangle\langle\psi_s| + (1-q)\tfrac{1}{4}I, \qquad (1)$$

where $|\psi_s\rangle$ is a singlet state, $I$ is the identity, and $q \in [0,1]$, lend themselves to a simple entanglement witness expressed as a sum of six projections involving only local measurements in one of three bases. By examining the partial transpose of $\rho_q$, one can show that the state is separable for $q \leqslant 1/3$ and entangled for $q > 1/3$. Furthermore, one can show that $\rho_q$ is classically correlated for $0 < q \leqslant 1/3$. The extreme values of $q$, then, correspond to a maximally mixed ($q = 0$) and maximally entangled ($q = 1$) state.

The simplicity of the Werner states and their associated entanglement witness lends itself well to experimental measurement. By using polarization-entangled photons, Barbieri *et al.* were able to prepare $\rho_q$ and measure $W$ for a variety of settings, thereby demonstrating entanglement in their system [14]. In terms of the horizontal and vertical polarization

---

[*]blacour@arlut.utexas.edu

modes (one-qubit states $|H\rangle$ and $|V\rangle$, respectively), the singlet state may be written as

$$|\psi_s\rangle = \tfrac{1}{\sqrt{2}}[|HV\rangle - |VH\rangle], \tag{2}$$

and the entanglement witness takes the form

$$W = \tfrac{1}{2}[|HH\rangle\langle HH| + |VV\rangle\langle VV| + |DD\rangle\langle DD| \\ + |AA\rangle\langle AA| - |LR\rangle\langle LR| - |RL\rangle\langle RL|], \tag{3}$$

where $|D\rangle = [|H\rangle + |V\rangle]/\sqrt{2}$, $|A\rangle = [|H\rangle - |V\rangle]/\sqrt{2}$, $|L\rangle = [|H\rangle + i|V\rangle]/\sqrt{2}$, and $|R\rangle = [|H\rangle - i|V\rangle]/\sqrt{2}$ represent the diagonal, antidiagonal, left-circular, and right-circular polarization states. It can be shown that, for the above Werner states, $\text{Tr}[W\rho_q] = (1-3q)/4$, so $0 \leqslant \text{Tr}[W\rho_q] \leqslant 0.25$ for $0 \leqslant q \leqslant 1/3$ and $-0.5 \leqslant \text{Tr}[W\rho_q] < 0$ for $1/3 < q \leqslant 1$.

In our classical analog of this experiment, we propose to follow the experimental procedure of Ref. [14], albeit using a classical (i.e., hidden variable) state in lieu of a quantum optical system. The details of state preparation and measurement are therefore radically different, but the types of measurements and the data analysis protocol are the same. A negative value of the measured entanglement witness so obtained will thereby be deemed indicative of entanglement. As we will see, the postselection of coincident-detection events, which is part of the analysis protocol, gives rise to contextuality. This will be of critical importance for obtaining a negative witness value.

## II. CLASSICAL REPRESENTATION

Our classical representation of the quantum state will consist of a random, complex-valued vector $\boldsymbol{a}$ of four components $a_{00}, a_{01}, a_{10}, a_{11}$ corresponding to the four standard basis states $|HH\rangle, |HV\rangle, |VH\rangle, |VV\rangle$, respectively. In the model, amplitude-threshold crossings correspond to measurements. The randomness associated with $\boldsymbol{a}$ will come from two sources: first, in state preparation, to model mixed states and, second, in an additive noise term, to model quantum statistics. The details of how these random values are drawn, as well as the process of measurement, are described in the following sections.

### A. Mixed states

A general mixed state takes the form of a convex sum of pure states, written as

$$\rho = \sum_i p_i |\psi_i\rangle\langle\psi_i|, \tag{4}$$

where $p_i \geqslant 0$ and $\sum_i p_i = 1$. Thus, generating a classical representation of the mixed state $\rho$ may be viewed as a straightforward matter of selecting, with each state preparation, a random pure state in accordance with the above probability distribution.

For the Werner state of Eq. (1), this process would consist of randomly choosing either $|\psi_s\rangle$, with probability $q$, or else choosing one of the four basis vectors $|HH\rangle, |HV\rangle, |VH\rangle$, or $|VV\rangle$ with equal probability. If we denote by $\tilde{\boldsymbol{\alpha}}$ the components of the randomly drawn state in the standard basis, we have

$$\tilde{\boldsymbol{\alpha}} = \begin{cases} \boldsymbol{\alpha}_s & \text{with probability } q \\ [1,0,0,0]^\mathsf{T} & \text{with probability } (1-q)/4 \\ [0,1,0,0]^\mathsf{T} & \text{with probability } (1-q)/4 \\ [0,0,1,0]^\mathsf{T} & \text{with probability } (1-q)/4 \\ [0,0,0,1]^\mathsf{T} & \text{with probability } (1-q)/4, \end{cases} \tag{5}$$

where $\boldsymbol{\alpha}_s = [0,1,-1,0]^\mathsf{T}/\sqrt{2}$. Of course, any two-qubit orthonormal basis would do (or, for that matter, any resolution of the identity, whether it is orthogonal or not).

Note that the density matrix, in the standard basis, corresponding to $\tilde{\boldsymbol{\alpha}}$ is given by

$$\boldsymbol{\rho}_q = E[\tilde{\boldsymbol{\alpha}}\tilde{\boldsymbol{\alpha}}^\dagger] = q\boldsymbol{\alpha}_s\boldsymbol{\alpha}_s^\dagger + (1-q)\tfrac{1}{4}\mathsf{I}, \tag{6}$$

where $\mathsf{I}$ is the identity matrix. This suggests that other choices for $\tilde{\boldsymbol{\alpha}}$ may adequately represent the Werner state as well. For example, a random vector of the form

$$\tilde{\boldsymbol{\alpha}} = \sqrt{q}\,\boldsymbol{\alpha}_s + \sqrt{1-q}\,\boldsymbol{w}, \tag{7}$$

where $\boldsymbol{w}$ is a zero-mean complex Gaussian random vector with covariance $\tfrac{1}{4}\mathsf{I}$, yields the same density matrix as that of Eq. (6). Note that $\tilde{\boldsymbol{\alpha}}$, as defined by Eq. (7), is not strictly normalized to unit magnitude but, rather, only on average (i.e., $E[\|\tilde{\boldsymbol{\alpha}}\|^2] = 1$). Nevertheless, as we will show, it does appear to provide an empirically equivalent description of the mixed state.

### B. Modeling quantum statistics

The above prescription for representing the quantum state says nothing of measurement. In Ref. [5] it was shown that, by adding a random vector $\boldsymbol{v}$ to a particular pure state $\boldsymbol{\alpha}$, scaled by $s > 0$, and observing threshold crossings of the component magnitude, one is able to reproduce or approximate the quantum statistics of the Born rule. More generally, replacing $\boldsymbol{\alpha}$ with $\tilde{\boldsymbol{\alpha}}$ for a mixed state, we have the random vector

$$\boldsymbol{a} = s\tilde{\boldsymbol{\alpha}} + \boldsymbol{v}, \tag{8}$$

where $\boldsymbol{v}$ is taken to be statistically independent of $\tilde{\boldsymbol{\alpha}}$. We shall refer to $\boldsymbol{v}$ as the *quantum noise* to distinguish it from the randomness in state preparation associated with $\tilde{\boldsymbol{\alpha}}$. The particular form of this model is motivated by stochastic electrodynamics [15], wherein the quantum noise term corresponds to the pertinent modes of a real (vice virtual), albeit random, vacuum field. Here we take it to simply be a mathematical model.

True photon detection in devices such as avalanche photodiodes arises from abrupt changes in current. Motivated by this, we shall say that a measurement of $\boldsymbol{a}$ results in the outcome $i$ if $|a_i| > \gamma$ and $|a_j| \leqslant \gamma$ for all $j \neq i$ for some threshold $\gamma > 0$. Measurements for which this is not the case, either because of multiple detections or a lack of detections, are discarded. In particular, if $\boldsymbol{v} = \sigma \boldsymbol{z}/\|\boldsymbol{z}\|$, where $\boldsymbol{z}$ is a standard complex Gaussian, $\sigma > 0$, $s = (\sqrt{2}-1)\sigma$, and $\gamma = \sigma$, then we are guaranteed that at most one threshold crossing will occur [5].

To measure in a basis other than the standard basis, one performs a unitary transformation corresponding to the desired measurement basis and then performs a threshold-detection measurement, as before. In the case of so-called *joint* measurements, the entire two-qubit state is transformed and measured in this manner. In the case of *local* measurements, each qubit is measured separately in a manner to be described below.

Thus, to perform a joint measurement of the projection $|HH\rangle\langle HH|$, for example, one examines $|a_{00}|$ and ascertains whether it is above the threshold $\gamma$. Similarly, to measure $|VV\rangle\langle VV|$ one examines $|a_{11}|$. For a joint measurement of $|DD\rangle\langle DD|$, say, one first transforms $\boldsymbol{a}$ to $\boldsymbol{a}' = (\mathsf{H}\otimes\mathsf{H})^\dagger\boldsymbol{a}$, where $\mathsf{H}$ is the unitary Hadamard matrix, and examines whether $|a'_{00}| > \gamma$. Finally, to measure $|LR\rangle\langle LR|$ jointly, we transform $\boldsymbol{a}$ to $\boldsymbol{a}' = (\mathsf{V}\otimes\mathsf{V})^\dagger\boldsymbol{a}$, where

$$\mathsf{V} = \frac{1}{\sqrt{2}}\begin{bmatrix} 1 & 1 \\ i & -i \end{bmatrix}, \tag{9}$$

and examine whether $|a'_{01}| > \gamma$. Other observables may be measured in a similar fashion.

To perform a local measurement of $|LR\rangle\langle LR|$, say, we first apply separate unitary transformations to qubits A and B, obtaining $\boldsymbol{a}' = (\mathsf{V}^\dagger\otimes\mathsf{I})\boldsymbol{a}$ and $\boldsymbol{b}' = (\mathsf{I}\otimes\mathsf{V}^\dagger)\boldsymbol{a}$, respectively. Next, we apply the projection matrix

$$\mathsf{P}_0^{(A)} = [1,0,0,0]^\mathsf{T}[1,0,0,0] + [0,1,0,0]^\mathsf{T}[0,1,0,0] \tag{10}$$

to $\boldsymbol{a}'$, obtaining $\boldsymbol{a}'' = \mathsf{P}_0^{(A)}\boldsymbol{a}'$, and the projection matrix

$$\mathsf{P}_1^{(B)} = [0,1,0,0]^\mathsf{T}[0,1,0,0] + [0,0,0,1]^\mathsf{T}[0,0,0,1] \tag{11}$$

to $\boldsymbol{b}'$, obtaining $\boldsymbol{b}'' = \mathsf{P}_1^{(B)}\boldsymbol{b}'$. Finally, we examine whether we have both $\|\boldsymbol{a}''\|^2 = |a'_{00}|^2 + |a'_{01}|^2 > \gamma^2$ and $\|\boldsymbol{b}''\|^2 = |b'_{01}|^2 + |b'_{11}|^2 > \gamma^2$. If this is so, then we say we have a coincident detection. Local measurements of other observables may be performed in a similar manner.

## III. EXPERIMENT SIMULATION

The notional classical experiment consists of a centralized source for generating random realizations of the vector $\boldsymbol{a}$ according to the procedures described in the previous section and for a given value of $q \in [0,1]$.

For local measurements, a copy of $\boldsymbol{a}$ is sent to separate devices to measure qubits A and B. Note that the same classical state is sent to each measuring device, but each device would measure a different qubit. Each device is capable of locally and independently selecting one of six possible measurements $(H, V, D, A, L,$ or $R)$. Projective threshold-crossing measurements are then performed and recorded. Each device measures the presence or absence of a threshold crossing, and coincident detections between qubits A and B are noted upon postanalysis.

Joint measurements are performed in a similar manner but require only one measurement device. The device is selectable between the six different projections in the definition of $W$ and indicates the presence or absence of the selected two-qubit state. So, a coincident detection of, say, $|HH\rangle\langle HH|$ merely indicates that $|a_{HH}|$ was found to fall above the threshold $\gamma$. If no threshold crossing is obtained, this fact is noted but the sample is discarded later during postanalysis.

We performed a numerical simulation of the above notional experiment, which we shall now describe. For the purposes of this study, we assume that the devices are able to faithfully encode the quantum state, as represented by the random vector $\boldsymbol{a}$, and reproduce the necessary unitary and projective transformations. We produced $N = 2^{20}$ realizations of $\boldsymbol{a}$ and counted the number of coincident detections for each of the six projection terms in $W$. Let $C_{HH}$ denote the number of coincident counts for $|HH\rangle\langle HH|$, $C_{VV}$ the number for $|VV\rangle\langle VV|$, $C_{DD}$ for $|DD\rangle\langle DD|$, and so on. We also measured $C_{HV}$ and $C_{VH}$, the coincidence counts for measurements of $|HV\rangle\langle HV|$ and $|VH\rangle\langle VH|$, respectively. Both local and joint measurements were performed.

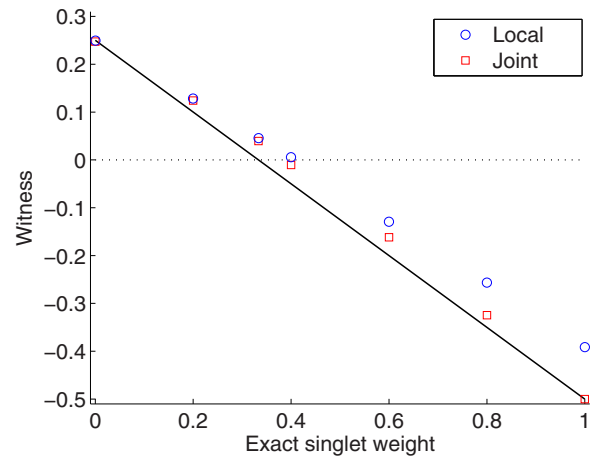

FIG. 1. (Color online) Plot of the estimated entanglement witness $W_{\text{est}}$ versus the exact singlet weight parameter $q$ using the mixture model of Eq. (5).

Generally, of the $N$ realizations, about 12% (8%) resulted in local (joint) coincident detections. This is comparable to the roughly $10^5$ coincident detections used in the experiment of Ref. [14]. Using the observed coincident counts, the entanglement witness was estimated by using the formula

$$W_{\text{est}} = \frac{1}{2}\frac{C_{HH} + C_{VV} + C_{DD} + C_{AA} - C_{LR} - C_{RL}}{C_{HH} + C_{HV} + C_{VH} + C_{VV}}, \tag{12}$$

where, following Ref. [14], we have normalized by the total coincident counts in the standard basis.

In Fig. 1 we plot the measured entanglement witness ($W_{\text{est}}$) versus the known value of the singlet weight parameter $q$ for values $q \in \{0, 0.2, 1/3, 0.4, 0.6, 0.8, 1.0\}$. In this example, $\tilde{\boldsymbol{\alpha}}$ was modeled according to Eq. (5). We find perfect agreement for the maximally mixed case ($q = 0$) and a general linear trend showing entanglement for $q > 0.4$. Perfect agreement is also found for the maximally entangled case ($q = 1$) when joint measurements are performed. In general, though, $W_{\text{est}}$ tends to fall slightly above the quantum prediction, with the joint measurement scheme showing somewhat better agreement than the local scheme.

We next consider modeling $\tilde{\boldsymbol{\alpha}}$ according to Eq. (7), with the results shown in Fig. 2. The estimates are qualitatively similar, with a general trend showing entanglement for large values of $q$. In this case, we note that joint measurements tend to give lower (more negative) witness values than the quantum prediction, while local measurements tend to give values that are slightly higher. Modifying the detection threshold tends to move the curves up (for $\gamma < \sigma$) or down (for $\gamma > \sigma$), with $\gamma = 1.05\sigma$ (local) and $\gamma = 0.95\sigma$ (joint) giving good agreement. The results in Fig. 2, however, are for $\gamma = \sigma$.

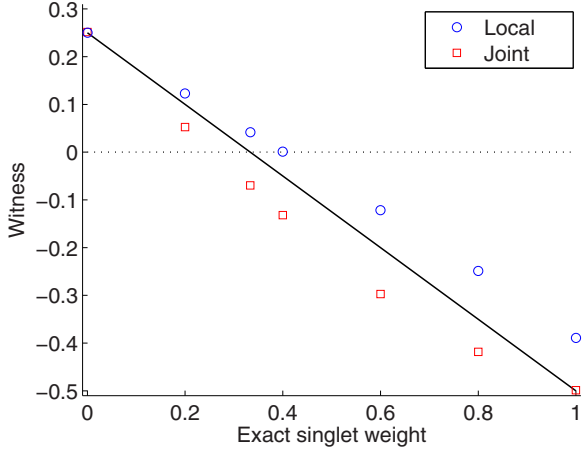

FIG. 2. (Color online) Plot of the estimated entanglement witness $W_{est}$ versus the exact singlet weight parameter $q$ using the mixture model of Eq. (7).

Now, unlike in the experiment of Ref. [14], we do have complete control and knowledge of the singlet weighting parameter $q$. Nevertheless, by following the same analysis protocol we may also estimate this parameter as follows:

First, we compute

$$R_{est} = \frac{1}{4}\left[\frac{C_{HH}}{C_{VH}} + \frac{C_{VV}}{C_{HV}} + \frac{C_{HH}}{C_{HV}} + \frac{C_{VV}}{C_{VH}}\right]. \quad (13)$$

Then, using the fact that the diagonal matrix elements of $\rho_q$ are $\langle HH|\rho_q|HH\rangle = \langle VV|\rho_q|VV\rangle = (1-q)/4$ and $\langle HV|\rho_q|HV\rangle = \langle VH|\rho_q|VH\rangle = (1+q)/4$, the parameter $q$ may be estimated by $q_{est} = (1 - R_{est})/(1 + R_{est})$.

Using this approach, we plot $W_{est}$ against the estimated singlet weight ($q_{est}$), as was done in Ref. [14]. These results are plotted in Fig. 3. Surprisingly, we find nearly perfect agreement with the quantum predictions, for both local and joint measurements. The results shown in Fig. 3 are for $\tilde{\alpha}$ modeled according to Eq. (5), but quite similar results were found for the Gaussian model of Eq. (7) as well. The corresponding counts are summarized in Table I. As shown in the Appendix, this agreement is well within the 95% confidence interval for statistical uncertainty. On the other hand, agreement in Figs. 1 and 2, which use the known values of $q$, would not be expected to improve with additional samples. Nevertheless, as we shall see in the next section, a change of threshold can give arbitrarily good agreement with the quantum predictions, even when $q$ is known.

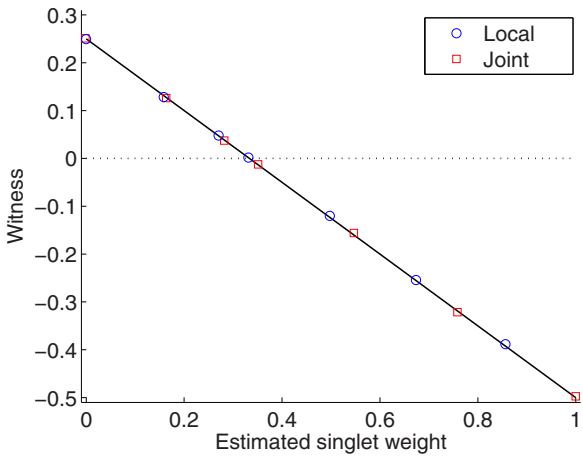

FIG. 3. (Color online) Plot of the estimated entanglement witness $W_{est}$ versus the estimated singlet weight parameter $q_{est}$ using the mixture model of Eq. (5). (Compare to Fig. 2 of Ref. [14]).

TABLE I. Table of measured counts for local measurements using the mixture model of Eq. (5), as illustrated in Fig. 3.

| $q$ | 0 | 0.2 | 1/3 | 0.4 | 0.6 | 0.8 | 1 |
|---|---|---|---|---|---|---|---|
| $C_{HH}$ | 31180 | 25796 | 22353 | 20297 | 14863 | 9660 | 4194 |
| $C_{HV}$ | 31287 | 35788 | 38688 | 40113 | 45075 | 49742 | 54250 |
| $C_{VH}$ | 31153 | 35668 | 39088 | 39938 | 45194 | 49572 | 53914 |
| $C_{VV}$ | 31360 | 25931 | 22327 | 20566 | 15178 | 9739 | 4212 |
| $C_{DD}$ | 29688 | 24406 | 21148 | 19400 | 14577 | 9416 | 4257 |
| $C_{AA}$ | 29735 | 24479 | 21291 | 19451 | 14448 | 9201 | 4164 |
| $C_{LR}$ | 29553 | 34590 | 37902 | 39275 | 44099 | 48880 | 53887 |
| $C_{RL}$ | 29525 | 34651 | 37883 | 39192 | 44248 | 49138 | 53797 |
| $W_{est}$ | 0.2516 | 0.1273 | 0.0463 | 0.0052 | −0.1217 | −0.2527 | −0.3897 |
| $q_{est}$ | −0.0008 | 0.1602 | 0.2703 | 0.3241 | 0.5006 | 0.6732 | 0.8558 |

## IV. DETECTOR EFFICIENCY

Conditioning on coincident counts is necessitated by the fact that not every state realization results in a measurement. This is true for quantum systems as well as the classical model considered here. This, then, leads to the concept of detector efficiency.

Notionally, detector efficiency would be defined as the ratio of detected photons to incident photons. Since the latter cannot be measured experimentally, various surrogate definitions are used instead, often based on attendant assumptions regarding statistical independence which cannot themselves be verified. In these schemes, one uses the measurable coincident counts $C_{HH}, C_{HV}, C_{VH}, C_{VV}$ as well as the single-detection counts $C_{HN}, C_{VN}, C_{NH}, C_{NV}$.

What remains inaccessible is the number of nondetections, $C_{NN}$, or, equivalently, the total number of state realizations, $N$. Unlike a true quantum system, our classical model provides full access to the underlying details of the experiment, including the values of $N$ and $C_{NN}$. This allows us to compute, for both joint and local measurements, the *true* coincident-detection efficiency as

$$\varepsilon_{AB} = \frac{C_{HH} + C_{HV} + C_{VH} + C_{VV}}{N}. \quad (14)$$

Efficiency for a single detector is computed by including the total number of single-detection events on that detector. Thus, the true efficiencies for measurements of qubits A and B, respectively, are given by

$$\varepsilon_A = \varepsilon_{AB} + \frac{C_{HN} + C_{VN}}{N}, \quad (15a)$$

$$\varepsilon_B = \varepsilon_{AB} + \frac{C_{NH} + C_{NV}}{N}. \quad (15b)$$

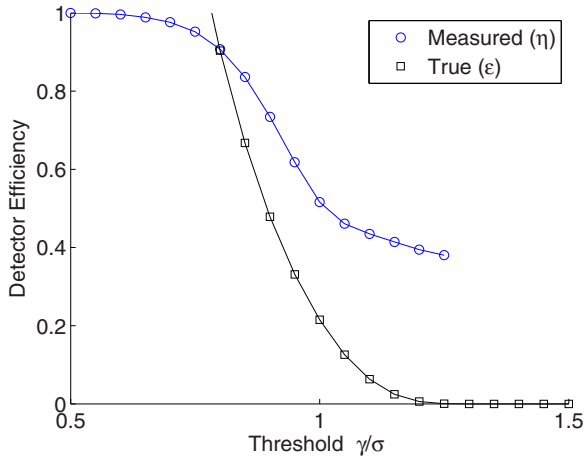

FIG. 4. (Color online) Plot of measured efficiency $\eta$ (blue circles) and true efficiency $\varepsilon$ (black squares) versus detection threshold $\gamma$ in units of $\sigma$ using local measurements on a singlet state.

Note that single-detector efficiencies make sense only for local measurements, because quantities such as $C_{HN}$ are not defined for joint measurements.

If the input states and detectors are physically identical, as is the case in our classical model, then it makes sense to define the true detector efficiency as simply the average of the two:

$$\varepsilon = \frac{\varepsilon_A + \varepsilon_B}{2}. \quad (16)$$

A plot of $\varepsilon$ versus the detection threshold $\gamma$ is shown in Fig. 4. Note that, for $\gamma \ll \sigma$, the efficiency becomes unbounded and therefore ill defined due to the presence of double detections (i.e., threshold crossings for both $H$ and $V$ detectors on a single qubit).

To measure the efficiency in the absence of $N$ or $C_{NN}$, one must make certain assumptions. Thus, one might suppose that the probability of obtaining $H$ on qubit A and $V$ on qubit B, say, which is estimated by $C_{HV}/N$, may be expressed as the "true" probability of obtaining $H$ and $V$, written $P_{HV}$ and predicted quantum mechanically to be $\langle HV|\rho_q|HV\rangle$, times the probability of a detection on qubit A and qubit B, written as $\eta_A \eta_B$. Thus, one assumes $C_{HV}/N \approx \eta_A \eta_B P_{HV}$. Similarly, $C_{HN}/N \approx \eta_A(1-\eta_B)(P_{HH}+P_{HV})$.

With these assumptions, we may deduce the following:

$$\eta_A = \varepsilon_{AB}/\varepsilon_B, \quad (17a)$$

$$\eta_B = \varepsilon_{AB}/\varepsilon_A. \quad (17b)$$

Since $N$ drops out in the ratio, both $\eta_A$ and $\eta_B$ are computable solely from measurements of coincident- and single-detection events. This is the manner in which detector efficiency is determined experimentally in a true quantum system. Finally, as symmetry warrants, the average of the two may be used to define the overall *measured* detector efficiency,

$$\eta = \frac{\eta_A + \eta_B}{2}. \quad (18)$$

A plot of $\eta$ versus the detection threshold $\gamma$ is shown in Fig. 4. Note that $\eta$ approaches 1 for $\gamma \ll \sigma$ and otherwise monotonically decreases with increasing threshold. For values of $\gamma$ greater than about $1.25\sigma$, there are no detections at all, so $\eta$ is undefined. We further note that, for $\gamma \gg 0.8\sigma$, the measured efficiency $\eta$ tends to be quite a bit larger than the true efficiency $\varepsilon$. For example, at $\gamma = \sigma$ we have $\eta = 0.52$ vice $\varepsilon = 0.22$. Typical efficiencies in coincident photon experiments tend to be around 0.3 at most [16], although efficiencies over 0.7 have been observed in supercooled detectors [17], so a value of 0.52 is actually quite good.

Implicit in the definition of the measured efficiency is the assumption that local detection events are independent when conditioned on a given quantum state. Thus, we write $\eta_A \eta_B$ as the probability of coincident detections and $\eta_A(1-\eta_B)$, say, as the probability of a detection only on qubit A. This assumption would indeed be reasonable in the naively realistic view that a quantum optic event such as the coincident detection of the state $|HV\rangle$ corresponds to an $H$ photon incident upon detector A and a $V$ photon incident upon detector B, the detection of each being determined by independent random variables intrinsic to each detector.

The present model provides an alternative to naive realism. In this view, all randomness originates in the initial signal, as represented by a particular realization of the complex vector *a*. Whether a detection event occurs or not is determined solely by *a* and not by any additional randomness associated with the detectors. Coincident detections are therefore correlated not only by virtue of correlations in the prepared state $\rho_q$ but also from correlations arising from the quantum noise term *v*. For this reason, we do not expect the coincident detector efficiency $\varepsilon_{AB}$ to factor into the product $\varepsilon_A \varepsilon_B$.

In fact, the true coincident efficiency $\varepsilon_{AB}$ and single-detector efficiencies $\varepsilon_A$ and $\varepsilon_B$ are related to the measured single-detector efficiencies $\eta_A$ and $\eta_B$ by $\varepsilon_{AB}^2/(\varepsilon_A \varepsilon_B) = \eta_A \eta_B$. To the extent that the single-detector efficiencies are equal, we furthermore have that $\varepsilon_{AB} = \varepsilon \eta$. Thus, the degree to which $\varepsilon$ and $\eta$ differ is indicative of the statistical dependence between local detection events. This, in turn, depends upon the detection threshold.

As we have now seen, detector efficiency can be made arbitrarily high by lowering the threshold. On the other hand, a lower threshold can also lead to undesirable effects such as double detections. This naturally leads one to question the impact of the threshold on measurements of the entanglement witness. In Fig. 5 we plot the estimated entanglement witness $W_{\text{est}}$ versus the threshold $\gamma$ for local measurements of the singlet state. We find that we approach the ideal witness value of $-0.5$ for $\gamma$ slightly above $\sigma$ and still remain entangled (i.e., $W_{\text{est}} < 0$) for $\gamma$ greater than about $0.88\sigma$. Thus, raising the threshold may be viewed as a mechanism for refining the level of entanglement, albeit at the cost of fewer detection events and an overall lower yield.

## V. DISCUSSION

It is important to understand that the results presented here rely fundamentally on the use of coincident detections to compute the estimated entanglement witness. If one normalizes by the actual number of realizations, $N$, instead of the total number of coincident counts, $C_{HH} + C_{HV} + C_{VH} + C_{VV}$, in the computation of $W_{est}$, no entanglement is observed (i.e., $W_{est} \geq 0$).

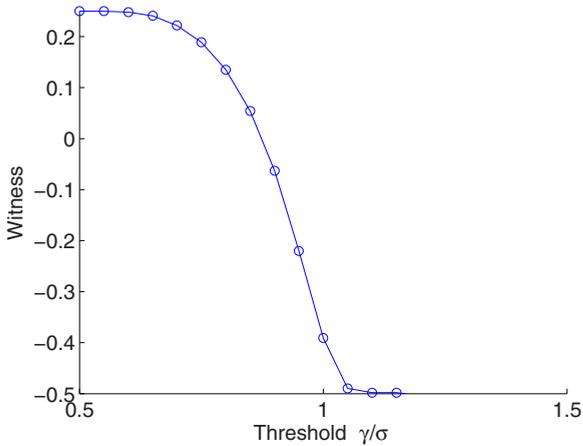

FIG. 5. (Color online) Plot of the entanglement witness $W_{est}$ versus detection threshold $\gamma$ in units of $\sigma$ using local measurements on a singlet state.

We may therefore infer the presence of entanglement no more strongly than could be inferred in the experiment of Ref. [14]. This is not to be construed as criticism of that experiment, though. The process of photon detection is inherently limited in efficiency, and the true number of photons impinging upon a detector, if that concept has any meaning, cannot be known. We do claim, however, that qualitatively similar results, in particular evidence of entanglement, can be obtained by using a classical system and a similar experimental procedure and data analysis protocol.

A larger question is whether the mathematical model described here could provide a physical explanation of the photon-polarization experiment itself. The present model was inspired by the work of Marshall and Santos on stochastic optics, a classical theory intended to explain quantum optics phenomena [18]. In that view, the random vector $\nu$ represents contributions from the vacuum zero-point field (ZPF), taken to be similar in form to that of quantum electrodynamics (QED) but assumed to be real rather than virtual. The threshold exceedance condition, then, merely reflects the response of a nonlinear device, such as an avalanche photodiode, to the intensity of a coherent sum of the prepared signal, represented by $s\tilde{\alpha}$, and the quantum background noise, represented by $\nu$. Replicating the quantum state to measure qubits A and B is, in this view, equivalent to dividing the incoming emission with a prism-like pair of mirrors.

Finally, although a particular probability distribution for the noise term $\nu$ was used here, other distributions may be considered as well. Unnormalized complex Gaussians, for example, have been found to produce similar results with suitably high thresholds [5].

## VI. CONCLUSION

We have described a classical model of joint and local measurements of an entanglement witness. By considering only coincident detections a negative witness value, indicative of entanglement, can be achieved. In general, joint measurements give rise to more strongly negative witness values than local measurements, although these values can be adjusted up or down by varying the detection threshold. Good agreement with quantum theoretic predictions can be achieved by adjusting the thresholds for joint and local measurements separately. Interesting, when plotted against the estimated singlet parameter, both joint and local measurements provide excellent agreement with theory.

These results point to an interesting subtlety in certain experimental determinations of entanglement. Optical experiments to verify entanglement, such as the one described in Ref. [14], are complicated by the lack of perfect efficiency in coincident photon detection. As a consequence, classical models such as the one described here may be able to provide an equally valid description of the observed phenomena. The model also provides an interesting perspective on detector efficiency. While efficiency is generally regarded as a property that is intrinsic to the detector, the present model suggests an alternative view in which the variability resides in the photon itself, so to speak, rather than in the measuring device.

## ACKNOWLEDGMENTS

This work was support by the Office of Naval Research under Grant No. N00014-14-1-0323 and an Internal Research and Development grant from Applied Research Laboratories, The University of Texas at Austin. The authors would like to thank Mark Selover, Kyungsun Na, Gopalakrishnan Bhamathi, and James Troupe for their useful discussions and comments on the manuscript.

## APPENDIX

In this appendix, we examine the construction of confidence intervals for the estimates $W_{est}$ and $q_{est}$. For large $N$, we may assume that count frequencies $C_{HH}/N$, $C_{HV}/N$, etc. are Gaussian distributed with mean $p_{HH}$, $p_{HV}$, etc. and standard deviations $\sigma_{HH} = \sqrt{p_{HH}(1-p_{HH})/N}$, $\sigma_{HV} = \sqrt{p_{HV}(1-p_{HV})/N}$, and so on. If these counts are the result of separate state preparations, then we may furthermore assume that they are statistically independent.

A standard propagation of errors for $W_{est}$ would suggest an uncertainty of the form

$$\delta W = |W_{est}| \sqrt{\left(\frac{\sigma_X}{\mu_X}\right)^2 + \left(\frac{\sigma_Y}{\mu_Y}\right)^2}, \quad \text{(A1)}$$

where

$$X = \frac{C_{HH} + C_{VV} + C_{DD} + C_{AA} - C_{LR} - C_{RL}}{N}, \quad \text{(A2)}$$

$$Y = \frac{C_{HH} + C_{HV} + C_{VH} + C_{VV}}{N}, \quad \text{(A3)}$$

and $\mu_X$, $\sigma_X$ are the mean and standard deviation, respectively, of $X$ (and similarly for $Y$). In practice, the sampled counts would be used in place of the exact expectation values.

A similar propagation of errors for $q_{est}$ yields

$$\delta q = |q_{est}| \sqrt{\left(\frac{\delta R}{1-\bar{R}}\right)^2 + \left(\frac{\delta R}{1+\bar{R}}\right)^2}, \quad \text{(A4)}$$

where

$$\delta R = \frac{1}{4}\sqrt{(\delta r_1)^2 + (\delta r_2)^2 + (\delta r_3)^2 + (\delta r_4)^2}, \quad (A5)$$

and

$$\delta r_1 = \frac{p_{HH}}{p_{VH}}\sqrt{\left(\frac{\sigma_{HH}}{p_{HH}}\right)^2 + \left(\frac{\sigma_{VH}}{p_{VH}}\right)^2}, \quad (A6)$$

$$\delta r_2 = \frac{p_{VV}}{p_{HV}}\sqrt{\left(\frac{\sigma_{VV}}{p_{VV}}\right)^2 + \left(\frac{\sigma_{HV}}{p_{HV}}\right)^2}, \quad (A7)$$

$$\delta r_3 = \frac{p_{HH}}{p_{HV}}\sqrt{\left(\frac{\sigma_{HH}}{p_{HH}}\right)^2 + \left(\frac{\sigma_{HV}}{p_{HV}}\right)^2}, \quad (A8)$$

$$\delta r_4 = \frac{p_{VV}}{p_{VH}}\sqrt{\left(\frac{\sigma_{VV}}{p_{VV}}\right)^2 + \left(\frac{\sigma_{VH}}{p_{VH}}\right)^2}. \quad (A9)$$

For $N = 2^{20}$, typical values of $\delta W$ and $\delta q$ are found to be no more than about 0.002, with a 95% confidence interval corresponding to about twice this value.

[1] M. A. Nielsen and I. L. Chuang, *Quantum Computation and Quantum Information* (Cambridge University Press, Cambridge, 2000).

[2] R. J. C. Spreeuw, A classical analogy of entanglement, Found. Phys. **28**, 361 (1998).

[3] P. Chowdhury, A. S. Majumdar, and G. S. Agarwal, Nonlocal continuous-variable correlations and violation of Bell's inequality for light beams with topological singularities, Phys. Rev. A **88**, 013830 (2013).

[4] P. Ghose and A. Mukherjee, Novel states of classical light and noncontextuality, Adv. Sci., Eng. Med. **6**, 246 (2014).

[5] B. R. La Cour, A locally deterministic, detector-based model of quantum measurement, Found. Phys. **44**, 1059 (2014).

[6] P. W. Shor, Algorithms for quantum computation: Discrete logarithms and factoring, in *Proceedings of 35th Annual Symposium on Foundations of Computer Science, 1994* (IEEE Computer Society, Washington, DC, 1994), p. 124.

[7] L. K. Grover, Quantum mechanics helps in searching for a needle in a haystack, Phys. Rev. Lett. **79**, 325 (1997).

[8] D. Deutsch and R. Jozsa, Rapid solutions of problems by quantum computation, Proc. R. Soc. London, Ser. A **439**, 553 (1992).

[9] R. Horodecki, P. Horodecki, M. Horodecki, and K. Horodecki, Quantum entanglement, Rev. Mod. Phys. **81**, 865 (2009).

[10] J. S. Bell, On the Einstein Podolsky Rosen paradox, Physics **1**, 195 (1964).

[11] B. M. Terhal, Bell inequalities and the separability criterion, Phys. Lett. A **271**, 319 (2000).

[12] O. Gühne, P. Hyllus, D. Bruß, A. Ekert, M. Lewenstein, C. Macchiavello, and A. Sanpera, Detection of entanglement with few local measurements, Phys. Rev. A **66**, 062305 (2002).

[13] R. F. Werner, Quantum states with Einstein-Podolsky-Rosen correlations admitting a hidden-variable model, Phys. Rev. A **40**, 4277 (1989).

[14] M. Barbieri, F. De Martini, G. Di Nepi, P. Mataloni, G. M. D'Ariano, and C. Macchiavello, Detection of entanglement with polarized photons: Experimental realization of an entanglement witness, Phys. Rev. Lett. **91**, 227901 (2003).

[15] L. de la Peña and A. M. Cetto, *The Quantum Dice: An Introduction to Stochastic Electrodynamics* (Springer, Dordrecht, 1995).

[16] A. Afriat and F. Selleri, *The Einstein, Podolsky, and Rosen Paradox in Atomic, Nuclear, and Particle Physics* (Plenum Press, New York, 1999).

[17] M. Giustina *et al.*, Bell violation using entangled photons without the fair-sampling assumption, Nature (London) **497**, 227 (2013).

[18] T. Marshall and E. Santos, Stochastic optics: A reaffirmation of the wave nature of light, Found. Phys. **18**, 185 (1988).